\documentclass[submission, PhysLectNotes]{SciPost}

\usepackage{amsmath}
\usepackage{amssymb}
\usepackage{amsthm} 
\usepackage{graphicx}
\usepackage{color}
\usepackage{braket}
\usepackage{hyperref}

\theoremstyle{plain}
\newtheorem{theorem}{Theorem}[section]

\newtheorem{lemma}[theorem]{Lemma}
\newtheorem{corollary}[theorem]{Corollary}
\newtheorem{definition}[theorem]{Definition}

\def\be{\begin{equation}}
\def\ee{\end{equation}}

\renewcommand{\bar}{\overline}

\begin{document}

\begin{center}{\Large \textbf{
Complex Lies, Real Physics: The Role of Algebra Complexification
}}\end{center}

\begin{center}
Tanguy Marsault\textsuperscript{1},
Laurent Schoeffel\textsuperscript{1*}
\end{center}

\begin{center}
{\bf 1} CEA Saclay, Irfu/SPP, Gif-sur-Yvette, France
\\
* laurent.schoeffel@cea.fr 
\end{center}

\begin{center}
\today
\end{center}

\section*{Abstract}
{\bf
In physics, Lie groups represent the algebraic structure that describes symmetry transformations of a given system. Then, the descending Lie algebra of those groups are necessarily real. In most cases, the complexification of those Lie algebras is necessary in order to derive irreducible representations of the Lie algebra and subsequently of the symmetry group. In this paper, we give a precise definition of the concept and prove step by step an important result $\left(\mathfrak{g}^\mathbb{R}\right)_\mathbb{C} \simeq \mathfrak{g} \times \bar{\mathfrak{g}}$.
This result is used to determine the irreducible representations of the proper Lorentz group and thus the physical objects admissible when this symmetry is present. It is shown that finite representations of the proper Lorentz group are characterized by pairs of half-integers $(j_1,j_2)$, which determine unambiguously the physical object associated to the given representation. For example, the representation $(0,0)$ of dimension $1$ is called the scalar representation, it corresponds to the Higgs field, and $(\frac{1}{2},0) \oplus (0,\frac{1}{2})$ of dimension $4$ is called the Dirac spinor representation, it corresponds to matter particle called fermions. This means that the mathematical group structure determines the material content of the universe following this algebraic structure. %
}

\vspace{10pt}
\noindent\rule{\textwidth}{1pt}
\tableofcontents\thispagestyle{fancy}
\noindent\rule{\textwidth}{1pt}
\vspace{10pt}

\section*{Notations}

Let us introduce some notations that will be used in the following.
\begin{itemize}
    \item [-] $\mathbb{K}$ is either $\mathbb{R}$ or $\mathbb{C}$, the field of real or complex numbers.
    \item [-] $M_n(\mathbb{K})$ is the set of $n\times n$ matrices with coefficients in $\mathbb{K}$.
    \item [-] $[\cdot,\cdot]$ always denotes the Lie bracket of the considered Lie algebra.
    \item [-] The Lie algebra descending from a Lie group $G$ is denoted in fraktur font $\mathfrak{g}$\dots
    \item [-] $\text{id}$ denotes the identity map on a vector space and $I_n$ the identity matrix of size $n$.
    \item [-] $\mathrm{Tr}$ denotes the trace of a matrix and $\mathrm{det}$ its determinant.
    \item [-] $\dim_\mathbb{K}$ denotes the dimension of a vector space over the field $\mathbb{K}$.
    \item [-] Working with finite dimensional vector spaces, we use interchangeably the notations $V\times W$ and $V\oplus W$ to denote the direct sum of two real vector spaces $V$ and $W$. For complex Lie algebras, the notation $\times$ always denotes the cartesian product of Lie algebras. If $\mathfrak{g}$ and $\mathfrak{h}$ are complex Lie algebras, $\mathfrak{g}\times\mathfrak{h}$ is the complex Lie algebra with scalar multiplication defined by $$\alpha\cdot (X,Y) = (\alpha\cdot_\mathfrak{g} X,\alpha\cdot_\mathfrak{h} Y)$$ for $\alpha\in\mathbb{C}$ and $(X,Y)\in \mathfrak{g}\times\mathfrak{h}$ and the Lie bracket defined by $$[(X,Y),(X',Y')] = ([X,X'],[Y,Y'])$$ for $(X,Y),(X',Y')\in \mathfrak{g}\times\mathfrak{h}$.\\
      \item [-]  The symbol $\simeq$ is used to specify that two spaces are isomorphic. For example, for two Lie algebra $\mathfrak{a}$ and $\mathfrak{b}$, $\mathfrak{a} \simeq \mathfrak{b}$
      means that there exists an isomorphism $\phi : \mathfrak{a} \rightarrow \mathfrak{b}$.
\end{itemize}

\section{Introduction}

A fundamental idea in physics is the notion of symmetry. When a system—whether described by particles, coordinates, or differential equations—remains unchanged under a set of transformations, we say that it possesses a symmetry. The set of such transformations typically forms a group. When the transformations depend smoothly on continuous parameters, the symmetry group is a Lie group.

Lie groups are often complicated global objects, but their local structure near the identity can be captured in a simpler, linear way. This linear structure is what we call the Lie algebra of the group: it is the tangent space to the group manifold at the identity, together with a natural operation (the Lie bracket) that reflects the group’s structure.
See \cite{paulin} for a general pedagogical presentation.

Unless stated explicitly, we assume these groups to be of finite dimension and closed subgroups of $GL_n(\mathbb{\mathbb{K}})$. Some of the results derived here remain valid for any Lie algebra. 

We assume that the main definitions are known concerning Lie groups, Lie algebra and the major properties of such structures. 
In order to pose some notations, let us briefly recall the main groups and algebra that are used in this paper :\begin{equation*}
\begin{aligned}
    &GL(n,\mathbb{K}) = \{ A \in M_n(\mathbb{K}),\; \mathrm{det}(A)\neq 0 \},\\
    &SL(n,\mathbb{K}) = \{ A \in M_n(\mathbb{K}),\; \mathrm{det}(A)=1 \},\\
    &SO(n) = \{ R \in GL_n(\mathbb{R}),\; R^\intercal R = I_n, \;\mathrm{det}(R)=1 \},\\
&SU(n) = \{ U \in GL_n(\mathbb{C}), \; U^\dagger U = I_n,\; \mathrm{det}(U)=1 \}.
\end{aligned}
\end{equation*}
The  Lie algebra 
descending from a Lie subgroup of $GL_n(\mathbb{K})$ is defined as
\begin{equation}
    \label{def:liealgebra}    
    \mathfrak{g}=\{ A \in M_n(\mathbb{K}),\; e^{t A} \in G \ \ \ \forall  t \in \mathbb{R} \}
\end{equation}
This definition leads to the following classical Lie algebras,
\begin{equation*}
\begin{aligned}
    &\mathfrak{gl}_n(\mathbb{K}) = M_n(\mathbb{K}),\\
    &\mathfrak{sl}_n(\mathbb{K}) = \{ A \in M_n(\mathbb{K}),\; \mathrm{Tr}(A)=0 \},\\
    &\mathfrak{so}(n) = \{ A \in M_n(\mathbb{R}),\; A^\intercal = -A \},\\
    &\mathfrak{su}(n) = \{ A \in M_n(\mathbb{C}),\; A^\dagger = -A,\; \mathrm{Tr}(A)=0 \}.
\end{aligned}
\end{equation*}
The Lie bracket on these classical Lie algebras is defined as the commutator of matrices, namely
$$
\forall X,Y \in M_n(\mathbb{K}), \ \ \left[ X,Y \right] = XY-YX.
$$

Let us give more details about the group $SU(2)$ and its Lie algebra $\mathfrak{su}(2)$, which are used in the following.
A basis of $\mathfrak{su}(2)$ is given by the three  matrices, 
\be
\label{beta}
\beta_1 = \frac{1}{2}
\begin{pmatrix}
0 & i  \\
i & 0 
\end{pmatrix}
\ \
\beta_2 = \frac{1}{2}
\begin{pmatrix}
0 & -1 \\
1 & 0 
\end{pmatrix}
\ \
\beta_3 = \frac{1}{2}
\begin{pmatrix}
i & 0  \\
0 & -i 
\end{pmatrix},
\ee
with the Lie brackets
$$
\left[ \beta_1, \beta_2 \right] = \beta_3 \ \ \  \left[ \beta_2, \beta_3 \right] = \beta_1  \ \ \ \left[ \beta_3, \beta_1 \right] = \beta_2.
$$
$\mathfrak{su}(2)$ is a real Lie algebra isomorphic to $\mathfrak{so}(3)$, the (real) Lie algebra of $SO(3)$, for which one basis reads
$$
J_1=
\begin{pmatrix}
0 & 0 & 0 \\
0 & 0 & -1 \\
0 & 1 & 0
\end{pmatrix} 
\ \
J_2 = \begin{pmatrix}
0 & 0 & 1 \\
0 & 0 & 0 \\
-1 & 0  & 0
\end{pmatrix}
\ \
J_3 = \begin{pmatrix}
0 & -1 & 0 \\
1 & 0 & 0 \\
0 & 0  & 0
\end{pmatrix},
$$
with
$$
\left[ J_1, J_2 \right] = J_3 \ \ \  \left[ J_2, J_3 \right] = J_1  \ \ \ \left[ J_3, J_1 \right] = J_2.
$$
The isomorphism between $\mathfrak{su}(2)$ and $\mathfrak{so}(3)$ is ensured by the equality of their structure constants, in the basis written above. This isomorphism is \emph{extraordinary}, in the sense that it cannot be generalized to higher dimensions, and has extremely important consequences in physics.

It is important to point out that Lie algebra descending from Lie groups are by definition real Lie algebras, hence also real vector spaces, see
$\mathfrak{su}(2)$ and $\mathfrak{so}(3)$. This is emphasized by the real nature of $t$ in Eq. (\ref{def:liealgebra}). Let us stress that the coefficients of the matrices can still be complex numbers, but the Lie algebra structure only makes sense under scalar multiplication by real numbers.

Let us give another important example with
the Lie algebra of the Lie group $SL(n,\mathbb{C})$, namely $\mathfrak{sl}(n,\mathbb{C}) = \left\{X\in M(n,\mathbb{C})\;|\;{\mathrm{Tr}} X = 0\right\}$.
Descending from a Lie group, $\mathfrak{sl}(n,\mathbb{C})$ is a \emph{real} Lie algebra. However, it can be trivially extended with a complex Lie algebra structure, since $\mathrm{Tr} (X)=0$ implies $\mathrm{Tr} (iX)=0$.
In the following $\mathfrak{sl}(n,\mathbb{C})$ is always considered as a complex Lie algebra. When we need to consider it as a real Lie algebra, we  resolve to its scalar restriction, $\mathfrak{sl}(n,\mathbb{C})^\mathbb{R}$.
Let us remark that the \emph{complex} dimension of $\mathfrak{sl}(n,\mathbb{C})$ is $n^2-1$, while the \emph{real} dimension of $\mathfrak{sl}(n,\mathbb{C})^\mathbb{R}$ is $2(n^2-1)$. Let us consider the simple case $n=2$. A basis of $\mathfrak{sl}(2,\mathbb{C})$, of dimension $3$, if given by
$$
H=
\begin{pmatrix}
1 & 0  \\
0 & -1  
\end{pmatrix} 
\ \
X = \begin{pmatrix}
0 & 1 \\
0 & 0 
\end{pmatrix}
\ \
Y = \begin{pmatrix}
0 & 0 \\
1 & 0 
\end{pmatrix},
$$
with the Lie brackets
$$
\left[ H, X \right] = 2X \ \ \  \left[ H, Y \right] = -2Y  \ \ \ \left[ X,Y \right] = H.
$$
Let us write 
$$
\mathfrak{sl}(2,\mathbb{C})=\{X,Y,H\},
$$
which indicates that the Lie algebra $\mathfrak{sl}(2,\mathbb{C})$ is spanned by the basis matrices $X$, $Y$, $H$. Then, we have 
$$
\mathfrak{sl}(2,\mathbb{C})^\mathbb{R}=\{X,Y,H,iX,iY,iH\}.
$$
It is important to stress that $\mathfrak{sl}(2,\mathbb{C})^\mathbb{R}$ is not $\mathfrak{sl}(2,\mathbb{R})$. Indeed, it is easy to see that a basis for $\mathfrak{sl}(2,\mathbb{R})$ is given by the three matrices $X$,$Y$,$H$ above with the same Lie brackets but, this time, $\mathfrak{sl}(2,\mathbb{R})$ is a real Lie algebra. Note that the same reasoning goes for $\mathfrak{gl}(n,\mathbb{C})$ which can also be endowed with a complex Lie algebra structure. On the contrary, $\mathfrak{su}(2)$ cannot be endowed with a complex Lie algebra structure because the antihermitian property is not conserved under scalar multiplication by $i$.

In physics theory, especially quantum mechanics, it is essential to derive irreducible representations from Lie groups or their descending Lie algebra. The picture is the following. Assuming a symmetry of the system under a group $G$, the Hilbert space of the quantum theory, $\mathcal{H}$, must carry a representation of the group $G$, that is there exists an homomorphism \begin{equation*}
    \rho: G \rightarrow GL(\mathcal{H}).
\end{equation*}
The theory being invariant under $G$, the Hamiltonian operator $H$ must commute with $\rho$, that is
\begin{equation*}
    \forall g\in G,\quad \rho(g) H = H \rho(g).
\end{equation*}

Some remarkable results from representation theory, with minor assumptions on the structure of $G$, prove that $\rho$ can be decomposed as
\begin{equation*}
    \rho = \rho_1 \oplus \ldots \oplus\rho_n,
\end{equation*}
where, $\rho_i$ is irreducible. In turn, this means that the Hilbert space may be decomposed in a similar fashion
\begin{equation*}
    \mathcal{H} = \mathcal{H}_1 \oplus \ldots \oplus\mathcal{H}_n,
\end{equation*}
so that $\rho_i: G \rightarrow GL(\mathcal{H}_i)$. Finally, Schur's lemma states that in such cases, the restriction of $H$ to each $\mathcal{H}_i$ is proportional to the identity \begin{equation*}
    H_{|\mathcal{H}_i} = \lambda_i \text{id}_{\mathcal{H}_i}.
\end{equation*}
It is pedagogical to pause here and remark how group theory already did most of the work. It provided us with the eigenstates of the theory, only the eigenvalues are determined by the physics. This procedure is met in most fields of modern physics, solid state physics being a prominent example. In this paper, we will rather point towards particle physics and the Lorentz group.

This rather sparse picture emphasizes how important it is to determine irreducible representations of a group $G$. It is obviously easier said than done. Working with Lie groups, one often encounters the need to analyze their representations in a more manageable form, through their Lie algebras. Studying representations of Lie algebras gives great insights into the representation theory of the original group, sometimes fully revealing its structure. Often, in the determination of the irreducible representations of a Lie algebra, the process of complexification of the Lie algebra is necessary.

Our purpose in this paper is to make some precise statements and proofs on how real Lie algebra can be complexified. We point out some 
subtleties that are generally hidden in these operations. Finally, we give an example on how this can be applied in the practice of physics theory by considering the irreducible representations the Lie algebra of proper Lorentz transformation.

\section{Complexification of a real Lie algebra}
\begin{definition}
    \label{definition:complexification}
    Let $\mathfrak{g}$ be a real Lie algebra. 
    The {complexified Lie algebra} $\mathfrak{g}_\mathbb{C}$ is the real vector space $\mathfrak{g} \times \mathfrak{g}$, endowed with scalar multiplication by the imaginary unit $i$ defined by
$$
    i \cdot (X, Y) = (-Y, X),
$$
    for any $X, Y \in \mathfrak{g}$ and equipped with the Lie bracket given by
\be
\label{lb1}
    [(X, Y), (X', Y')] = \big([X, X'] - [Y, Y'],\; [X, Y'] + [Y, X'] \big).
\ee
\end{definition}
Then $\mathfrak{g}_\mathbb{C}$ has a Lie algebra structure, as the anti-symmetry and the Jacobi Identity of the Lie bracket follows immediately from the definition above. Let us give a few remarks and examples following this definition.
\begin{itemize}

\item[-] Let $\mathfrak{g}$ be a real Lie algebra. Consider the space 
        $$
        \tilde{\mathfrak{g}} = \mathfrak{g}\oplus i\mathfrak{g}
        $$
        equipped with the external product 
    \begin{eqnarray*}
    \mathbb{C} \times (\mathfrak{g} \oplus i \mathfrak{g}) &\rightarrow& (\mathfrak{g} \oplus i \mathfrak{g})  \\
        (z = x + i y, V = A + iB) &\mapsto& z V = (x A - y B) + i (x B + y A),
    \end{eqnarray*}
    and endowed with the Lie bracket, 
\be
\label{lb2}
     [X+iY,X'+iY'] = \left(  [X,X'] -  [Y,Y']\right) + i \left(  [Y,X'] +  [X,Y']\right).
\ee
    It can be checked that $\tilde{\mathfrak{g}}$ is a Lie algebra. \\
    Define the map from $\mathfrak{g}_\mathbb{C}$ to $\tilde{\mathfrak{g}}$
    $$
    \begin{array}{rlcl}
       \phi:&\mathfrak{g}_\mathbb{C}  &\rightarrow & \tilde{\mathfrak{g}}  \\
         &(X,Y)&\mapsto& X+iY.
    \end{array}
    $$
Using Eq. (\ref{lb1}) and (\ref{lb2}), we obtain  that   $\phi$ is an isomorphism of complex Lie algebras. Indeed, $\phi$ is a bijective map and
for $(X_1,Y_1)$ and $(X_2,Y_2)$ in  $\mathfrak{g}_\mathbb{C}=\mathfrak{g} \times \mathfrak{g}$, with $a,b \in \mathbb{R}$, we have
$$
\phi(a (X_1,Y_1)+b (X_2,Y_2)) = a \phi(X_1,Y_1) + b \phi(X_2,Y_2),
$$
\begin{eqnarray*}
\phi( [(X_1,Y_1),(X_2,Y_2)]) &=& \phi(\big([X_1, X_2] - [Y_1, Y_2],\; [X_1, Y_2] + [Y_1, X_2] \big)) \\
  &=& [X_1, X_2] - [Y_1, Y_2] + i \big( [X_1, Y_2] + [Y_1, X_2] \big) \\
  &=&  [X_1+iY_1,X_2+iY_2] \\
  &=& [\phi(X_1,Y_1),\phi(X_2,Y_2)].
\end{eqnarray*}

\noindent
This makes clear the denomination of \emph{complexification}. The above construction $\tilde{\mathfrak{g}} = \mathfrak{g}\oplus i\mathfrak{g}$
with the Lie bracket (\ref{lb2})
can also be taken as a definition of \emph{complexification}.

\item[-] Let $\mathfrak{g}$ and $\mathfrak{h}$ be 2 real Lie algebras, we have
$$
    (\mathfrak{g}\times\mathfrak{h})_\mathbb{C} \simeq \mathfrak{g}_\mathbb{C}\times \mathfrak{h}_\mathbb{C}
$$
    \end{itemize}

It is also possible  to go the other way around, that is start from a complex Lie algebra and define a real Lie algebra, this is called the scalar restriction of a complex Lie algebra.    
\begin{definition}
    Let $\mathfrak{g}$ be a complex Lie algebra. The scalar restriction of $\mathfrak{g}$ is the real vector space $\mathfrak{g}^\mathbb{R}$ with basis the real and imaginary parts of the basis of $\mathfrak{g}$, endowed with the Lie bracket of $\mathfrak{g}$. \\
    To be more precise, let $\left(X_1,\ldots,X_n\right)$ be a basis of $\mathfrak{g}$, then the basis of $\mathfrak{g}^\mathbb{R}$ is given by
$$
    \left(X_1,\ldots,X_n,\; i X_1,\ldots,i X_n\right).
$$
The scalar multiplication acts as $(x, A)\mapsto x A$ for $x\in\mathbb{R}$ and $A\in \mathfrak{g}^\mathbb{R}$ and the Lie bracket is the one of the complex Lie algebra $\mathfrak{g}$.
\end{definition}

It is natural to wonder whether these two operations are inverses of each other. The answer is no, this can be easily seen by looking at the dimensions of the vector spaces. Let $\mathfrak{g}$ be a complex Lie algebra. Its scalar restriction has dimension $$
\dim_\mathbb{R} \mathfrak{g}^\mathbb{R} = 2 \dim_\mathbb{C} \mathfrak{g}.
$$
The complexification does not change the dimension of the vector space, $$
\dim_\mathbb{C} \left(\mathfrak{g}^\mathbb{R}\right)_\mathbb{C} = \dim_\mathbb{R} \mathfrak{g}^\mathbb{R} = 2 \dim_\mathbb{C} \mathfrak{g}.
$$
Which proves directly that, $$
    \left(\mathfrak{g}^\mathbb{R}\right)_\mathbb{C} \neq \mathfrak{g}.
$$
However, we can still derive a relationship between the successive applications of the two operations and the original Lie algebra.
\begin{theorem}
\label{theorem:restrictionExtension}
Let $\mathfrak{g}$ be a complex Lie algebra, $\mathfrak{g}^\mathbb{R}$ its scalar restriction and $\left(\mathfrak{g}^\mathbb{R}\right)_\mathbb{C}$
the complexification of this scalar restriction, we have 
$$
    \left(\mathfrak{g}^\mathbb{R}\right)_\mathbb{C} \simeq \mathfrak{g} \times \bar{\mathfrak{g}},
$$
where the conjugate Lie algebra $\bar{\mathfrak{g}}$ of $\mathfrak{g}$ is 
the Lie algebra with the same vector space structure as $\mathfrak{g}$, the same Lie bracket but with the scalar multiplication conjugated, namely
$$
\forall \alpha\in\mathbb{C},\; \forall X\in\mathfrak{g}\quad \alpha \cdot_{\bar{\mathfrak{g}}} X = \bar{\alpha} \cdot_{\mathfrak{g}} X.
$$

\begin{proof}
By definition $\left(\mathfrak{g}^\mathbb{R}\right)_\mathbb{C}$ has the complex vector space structure of $\mathfrak{g}^\mathbb{R}\times \mathfrak{g}^\mathbb{R}$ with the action of $i$ defined by 
$$
    \begin{array}{rlcl}
         i:& \mathfrak{g}^\mathbb{R}\times \mathfrak{g}^\mathbb{R} &\rightarrow & \mathfrak{g}^\mathbb{R}\times \mathfrak{g}^\mathbb{R} \\
         & (X,Y)&\mapsto& (-Y,X),
    \end{array}
$$
and the Lie bracket defined as, 
$$
[(X,Y),(X^\prime,Y^\prime)] = ([X,X^\prime] - [Y,Y^\prime], [X,Y^\prime]+[Y,X^\prime]).
$$
This is to be compared with the structure of $\mathfrak{g}\times\bar{\mathfrak{g}}$ which has action of $i$ defined by$$
    \begin{array}{rlcl}
         i:& \mathfrak{g}\times\bar{\mathfrak{g}} &\rightarrow & \mathfrak{g}\times\bar{\mathfrak{g}} \\
         & (X,Y)&\mapsto& (i\cdot_\mathfrak{g} X, i\cdot_{\bar{\mathfrak{g}}} Y) = (iX,-iY),
    \end{array}
$$
and the Lie bracket reads, 
$$
[(X,Y),(X^\prime,Y^\prime)] = ([X,X^\prime],[Y,Y^\prime]).
$$ 

Define the map $\Theta$ as, 
$$
    \begin{array}{rccl}
         \Theta : &\mathfrak{g}^\mathbb{R}\times \mathfrak{g}^\mathbb{R} & \rightarrow &\mathfrak{g}\times\bar{\mathfrak{g}}\\
         &(X,Y) &\mapsto& (X+iY,X-iY)
    \end{array}
$$
We  show that $\Theta$ is a Lie algebra isomorphism. It can be checked that it is an invertible map whose inverse is 
$$
    \begin{array}{rccl}
         \Theta^{-1} : &\mathfrak{g}\times\bar{\mathfrak{g}} & \rightarrow &\mathfrak{g}^\mathbb{R}\times \mathfrak{g}^\mathbb{R}\\
         &(A,B) &\mapsto &\left(\frac{1}{2}(A+B),\frac{1}{2i}(A-B)\right)
    \end{array}
$$
The $\mathbb{R}$-linearity of $\Theta$ is obvious and only its commutation with $i$ needs to be checked in order to make it a complex vector space isomorphism
\begin{align*}
    \Theta(i\cdot(X,Y)) &= \Theta(-Y,X) = (-Y+iX,-Y-iX)\\
    &=\left(i\cdot_\mathfrak{g}(X+iY), i\cdot_{\bar{\mathfrak{g}}}(X-iY)\right)\\
    &=i\left(X+iY,X-iY\right)\\
    &= i \Theta(X,Y).
\end{align*}
This proves that $\Theta$ defines a $\mathbb{C}$-linear isomorphism of vector spaces. Let us check that it commutes with the Lie bracket to complete the proof 
\begin{align*}
    \Theta\left([(X,Y),(X^\prime,Y^\prime)]\right) &=  \Theta\left([X,X^\prime]-[Y,Y^\prime], [X,Y^\prime]+[Y,X^\prime]\right)\\
    &=\left([X,X^\prime]-[Y,Y^\prime]+i[X,Y^\prime]+i[Y,X^\prime],[X,X^\prime]-[Y,Y^\prime]-i[X,Y^\prime]-i[Y,X^\prime]\right)\\
    &=\left([X+iY,X^\prime+iY^\prime],[X-iY,X^\prime-iY^\prime]\right)\\
    &=[\Theta(X,Y),\Theta(X^\prime,Y^\prime)].
\end{align*}

An alternative proof can be done using the second definition of the complexification of a real Lie algebra, namely
$ \left(\mathfrak{g}^\mathbb{R}\right)_\mathbb{C} = \mathfrak{g}^\mathbb{R} \oplus i \mathfrak{g}^\mathbb{R}$.
Any element of $\left(\mathfrak{g}^\mathbb{R}\right)_\mathbb{C}$ can be written as $X+iY$ with 
$(X,Y) \in  \mathfrak{g}^\mathbb{R}\times \mathfrak{g}^\mathbb{R}$.

Let us define  a $\mathbb{R}$-linear map $J$ as
\begin{eqnarray*}
        J : \mathfrak{g}^\mathbb{R} &\rightarrow& \mathfrak{g}^\mathbb{R} \\
        X &\mapsto& J(X)
\end{eqnarray*}
such that 
$$
\forall X \in \mathfrak{g}^\mathbb{R}, \ \ J^2(X)=-X.
$$
This map can be extended easily as a $\mathbb{C}$-linear map acting on $\left(\mathfrak{g}^\mathbb{R}\right)_\mathbb{C}$ by
$$
\forall (X,Y) \in  \mathfrak{g}^\mathbb{R}\times \mathfrak{g}^\mathbb{R}, \ \   X+iY \in \left(\mathfrak{g}^\mathbb{R}\right)_\mathbb{C}, \ \ J_\mathbb{C}(X+iY) = J(X) + i J(Y).
$$
$J_\mathbb{C}$ defined in this way is obviously a $\mathbb{C}$-linear map such that :
\begin{eqnarray*}
        J_\mathbb{C} : \left(\mathfrak{g}^\mathbb{R}\right)_\mathbb{C} &\rightarrow& \left(\mathfrak{g}^\mathbb{R}\right)_\mathbb{C} \\
        V &\mapsto& J(V)
\end{eqnarray*}
always with 
$$
\forall V \in \left(\mathfrak{g}^\mathbb{R}\right)_\mathbb{C},\quad J_\mathbb{C}^2(V)=-V.
$$
As $J_\mathbb{C}^2 = -\mathrm{id}$, we have
$$
     \left(\mathfrak{g}^\mathbb{R}\right)_\mathbb{C}=\ker (J_\mathbb{C}^2+\mathrm{id}).
$$
Remark that the polynomial $P(X) = X^2 + 1$ splits completely over the field $\mathbb{C}$ as
$$
    P(X) = (X-i)(X+i),
$$
and the two factors are coprime. In turn, the kernel decomposition lemma states
$$
\left(\mathfrak{g}^\mathbb{R}\right)_\mathbb{C} = \ker P(J) = \ker (J_\mathbb{C}-i \;\mathrm{id}) \oplus \ker (J_\mathbb{C}+i\;\mathrm{id}).
$$
Here, we have trivially
$$
\ker (J_\mathbb{C}-i \;\mathrm{id}) \oplus \ker (J_\mathbb{C}+i\;\mathrm{id}) \simeq \ker (J_\mathbb{C}-i \;\mathrm{id}) \times \ker (J_\mathbb{C}+i\;\mathrm{id}).
$$
The advantage of the notation $\ker (J_\mathbb{C}-i \;\mathrm{id}) \times \ker (J_\mathbb{C}+i\;\mathrm{id})$ is that it naturally encodes the
scalar multiplication and Lie bracket structure when we turn to Lie algebra.

Let us denote $\mathfrak{g}_\pm = \ker (J_\mathbb{C} \mp i\;\mathrm{id})$, so that we can rewrite
$$
 \left(\mathfrak{g}^\mathbb{R}\right)_\mathbb{C}
 = 
\mathfrak{g}_+\times \mathfrak{g}_-.
$$
Explicitly, we have
$$
\mathfrak{g}_+ = \{V_+ \in  \left(\mathfrak{g}^\mathbb{R}\right)_\mathbb{C} / J_\mathbb{C}(V_+)=+iV_+ \}.
$$
$$
\mathfrak{g}_- = \{V_- \in  \left(\mathfrak{g}^\mathbb{R}\right)_\mathbb{C} / J_\mathbb{C}(V_-)=-iV_- \}.
$$
Remark that $\mathfrak{g}_+$ has the complex structure of $\mathfrak{g}$ and $\mathfrak{g}_-$ has the complex structure of $\bar{\mathfrak{g}}$. 
Noting that
$$
\dim \mathfrak{g}_\pm = \frac{1}{2} \dim \left(\mathfrak{g}^\mathbb{R}\right)_\mathbb{C} = \dim \mathfrak{g} = \dim\bar{\mathfrak{g}},
$$
the following Lie algebra identities are shown
$$
\mathfrak{g}_+ = \mathfrak{g}\quad \text{and}\quad \mathfrak{g}_- = \bar{\mathfrak{g}}
$$
Finally,
$$
    \left(\mathfrak{g}^\mathbb{R}\right)_\mathbb{C} \simeq  \mathfrak{g} \times \bar{\mathfrak{g}},
$$
which completes the proof. 
\end{proof} 
\end{theorem}

In physics we work mostly with simple or semisimple Lie algebras, for which the previous result greatly simplifies. Let us first prove two lemmas.

\begin{lemma}
\label{lemma:semisimpleReal}
Let $\mathfrak{g}$ be a semisimple complex Lie algebra, then there exists a basis of $\mathfrak{g}$ such that the structure constants are real.
\begin{proof}
    This comes from the very powerful result that we admit that if $\mathfrak{g}$ is a semisimple complex Lie algebra, then there exists $\mathfrak{h}$ a real form $\mathfrak{g}$, that is $\mathfrak{h}_\mathbb{C}\simeq \mathfrak{g}$. The structure constants of $\mathfrak{h}$ are therefore real, and in the basis realizing the isomorphism $\mathfrak{h}_\mathbb{C}\simeq \mathfrak{g}$, the structure constants of $\mathfrak{g}$ are also real.
\end{proof}
\end{lemma} 

\begin{lemma}
\label{lemma:conjIsomorphism}
Let $\mathfrak{g}$ be a complex Lie algebra. If there exists a basis of $\mathfrak{g}$ such that the structure constants are real, then $\mathfrak{g} \simeq \bar{\mathfrak{g}}$ as Lie algebras.
\begin{proof}
    Let us denote by $\left(X_1,\ldots,X_n\right)$ such a basis. The structure constants of $\mathfrak{g}$ in this basis are defined by
$$
        [X_i,X_j] = \sum_{k=1}^n c_{ijk} X_k,
$$ 
where $c_{ijk} \in \mathbb{R}$. 
    Define, 
\begin{equation*}
    \begin{array}{rlcl}
        \phi: &\mathfrak{g} &\longrightarrow& \bar{\mathfrak{g}} \\
        &\sum_i \lambda_i \cdot_{\mathfrak{g}} X_i &\longmapsto& \sum_i \lambda_i \cdot_{\bar{\mathfrak{g}}} X_i,
    \end{array}
\end{equation*}
$\phi$ is a linear map for the complex structure of $\mathfrak{g}$ and the conjugated complex structure of $\bar{\mathfrak{g}}$ and defines an isomorphism of vector spaces.\\
Let us now check that $\phi$ is a Lie algebra isomorphism. 

Let $Y,Z\in\mathfrak{g}$, $Y=\sum_i \lambda_i X_i$ and $Z = \sum_j \mu_j X_j$, then
\begin{align*}
    \phi([Y,Z]) &= \phi\left(\sum_{i,j} c_{ijk} \lambda_i \mu_j X_k\right) \\
    &= \sum_{i,j} c_{ijk} \bar{\lambda}_i \bar{\mu}_j X_k \\
    &= \sum_{i,j} \bar{\lambda}_i \bar{\mu}_j [X_i,X_j] \\
    &= \left[\sum_i \bar{\lambda}_i X_i,\sum_j \bar{\mu}_j X_j\right] \\
    &= \left[\phi(Y),\phi(Z)\right],
\end{align*}
which completes the proof.
\end{proof}
\end{lemma}

These two lemmas allow us to state the following corollary for semisimple Lie algebras.
\begin{corollary}
    Let $\mathfrak{g}$ be a semisimple Lie algebra, then
    \begin{equation*}
        \left(\mathfrak{g}^\mathbb{R}\right)_\mathbb{C} \simeq \mathfrak{g} \times \mathfrak{g}.
    \end{equation*}
    \begin{proof}
        It is a direct consequence of Theorem (\ref{theorem:restrictionExtension}), Lemma (\ref{lemma:semisimpleReal}) and Lemma (\ref{lemma:conjIsomorphism}).
    \end{proof}
\end{corollary}

\section{Application to $\mathfrak{sl}(2,\mathbb{C})$ and consequences}

The laws of physics are said to be invariant under Lorentz transformations. These transformations corresponds to the linear transformations of the four-dimensional Minkowski space $\mathbb{R}^{1,3}$ that preserve the Minkowski metric $\eta = \mathrm{diag}(-1,1,1,1)$. In physics, usually one restricts to the proper orthochronous Lorentz transformations, that is the Lorentz transformations that preserve the orientation of space and the direction of time. Mathematically, it is described by the set $SO^0(1,3)$,
$$
    SO^0(1,3) = \{ R \in GL_4(\mathbb{R}), R^\intercal \eta R = \eta, \quad \mathrm{det}(R)=1,\quad R_{00} > 0, \quad \eta = \mathrm{diag}(-1,1,1,1) \}.
$$
In quantum mechanics, the invariance under $SO^0(1,3)$ means that the Hilbert space $\mathcal{H}$ of physical states carries a representation of $SO^0(1,3)$. More explicitly, a representation of the Lie group $SO^0(1,3)$ consists of a Hilbert space $\mathcal{H}$ together with an homomorphism $\rho$ such that
$$
    \rho: SO^0(1,3) \rightarrow GL(\mathcal{H}).
$$
Saying that $SO^0(1,3)$ is a symmetry of the physics system means that the Hamiltonian of the physics system is invariant under this representation,
$$
    \forall R \in SO^0(1,3),\quad H \rho(R) = \rho(R)H.
$$
A particle is often defined as an irreducible representation of finite dimension of $SO^0(1,3)$ (actually rather of the Poincaré group as is carefully explained in Appendix \ref{appendix6}), that is a pair $(\rho_i, \mathcal{H}_i)$, where $\mathcal{H}_i \subset \mathcal{H}$ is a subspace of $\mathcal{H}$ such that $\rho_i(R) \mathcal{H}_i = \mathcal{H}_i$ for all $R \in SO^0(1,3)$ and $\rho_i$ is irreducible, that is there is no non-trivial subspace of $\mathcal{H}_i$ invariant under $\rho_i(R)$. To understand what kind of particles can evolve in the universe, we need to determine the irreducible representations of $SO^0(1,3)$.
Indeed, by kind of particles, we mean elements of the Hilbert space $\mathcal{H}$, on which the transformation $\rho(R) \in GL(\mathcal{H})$ is acting.

A general result is that any irreducible representation of the Lie algebra $\mathfrak{g}$ of a Lie group $G$ can be lifted to an irreducible representation of the group $G$ itself through exponentiation. This means that we can already determine \emph{most} of the irreducible representations of $SO^0(1,3)$ by studying the irreducible representations of its Lie algebra $\mathfrak{so}(1,3)$. The question of whether all of the representations of $SO^0(1,3)$ can be lifted from $\mathfrak{so}(1,3)$ is a subtle one. The answer is no but proving it would take us far away from the scope of this paper. 

To determine the irreducible representations of $\mathfrak{so}(1,3)$, one can resolve to its complexification $\mathfrak{so}(1,3)_\mathbb{C}$. 
Indeed, there is a bijective correspondance between the equivalence classes of irreducible representations of 
a real Lie algebra and its complexification. 
This follows from the isomorphism
\be
\label{hom}
\mathrm{Hom}_\mathbb{R} ( \mathfrak{g}^\mathbb{R}, \mathfrak{h}) \simeq \mathrm{Hom}_\mathbb{C} ( \left(\mathfrak{g}^\mathbb{R}\right)_\mathbb{C}, \mathfrak{h}),
\ee
given $\mathfrak{g}$ and $\mathfrak{h}$ complex Lie algebras and
where $\mathrm{Hom}_\mathbb{R} (.)$ is the vector space of all real-linear Lie algebra homomorphisms and $\mathrm{Hom}_\mathbb{C} (.)$
the vector space of all complex-linear Lie algebra homomorphisms. 
The proof of Eq. (\ref{hom}) is simple. 
We define a map from the set of real homomorphisms to the set of complex homomorphisms as
$$
\Phi : \phi \in \mathrm{Hom}_\mathbb{R} ( \mathfrak{g}^\mathbb{R}, \mathfrak{h}) \rightarrow 
\psi=\Phi(\phi), 
$$
such that for any $(X,Y) \in \mathfrak{g}^\mathbb{R} \times  \mathfrak{g}^\mathbb{R} $, $\psi$ is extending $\phi$ $\mathbb{C}$-linearly
$$
\psi(X+iY)=\phi(X)+i\phi(Y).
$$
This is immediate to verify that the map $\psi$ is $\mathbb{C}$-linear and preserves the Lie bracket, therefore 
$\psi \in \mathrm{Hom}_\mathbb{C} ( \left(\mathfrak{g}^\mathbb{R}\right)_\mathbb{C}, \mathfrak{h})$ and $\Phi$ is a well defined map.
The map $\Psi$ from $\mathrm{Hom}_\mathbb{C} ( \left(\mathfrak{g}^\mathbb{R}\right)_\mathbb{C}, \mathfrak{h})$ to
$\mathrm{Hom}_\mathbb{R} ( \mathfrak{g}^\mathbb{R}, \mathfrak{h})$ is defined such that,
given a complex homomorphism, its image is obtained by restricting its domain to the real part of the complexification.
It is immediate to verify that
$\Psi$ is also a well defined map and $\Psi \circ \Phi = \Phi \circ \Psi = \text{id}$, which concludes.

The following result will be useful. Its proof is given in the appendix 5.

\begin{lemma}
\label{theorem:so13sl2C}
$$
\mathfrak{so}(1,3)_\mathbb{C} \simeq \left(\mathfrak{sl}(2,\mathbb{C})^\mathbb{R}\right)_\mathbb{C}
$$
\end{lemma}

In fact, we have a stronger result than
Lemma (\ref{theorem:so13sl2C}) as the following statement is correct
$$
\mathfrak{sl}(2,\mathbb{C})^\mathbb{R} \simeq \mathfrak{so}(1,3).
$$
However, the proof of this more general result is quite involved and would take us out of the scope of this paper. 
Moreover, as already mentioned, thanks to the 
correspondance between the equivalent classes of irreducible representations of 
a real Lie algebra and its complexification, we only need Lemma (\ref{theorem:so13sl2C}) to conclude.

Let us now apply the result of Theorem (\ref{theorem:restrictionExtension}) to the case of $\mathfrak{sl}(2,\mathbb{C})$, which is a semisimple Lie algebra.
\be
\label{sl2a}
    \left(\mathfrak{sl}(2,\mathbb{C})^\mathbb{R}\right)_\mathbb{C}  \simeq \mathfrak{sl}(2,\mathbb{C}) \times {\mathfrak{sl}(2,\mathbb{C})}.
\ee
There only remains to determine the irreducible representations of the complex Lie algebra $\mathfrak{sl}(2,\mathbb{C})$ in order to derive the irreducible representations of $\left(\mathfrak{sl}(2,\mathbb{C})^\mathbb{R}\right)_\mathbb{C}$. This is easily done using the following lemma.
\begin{lemma}
\be
\label{sl2b}
{\mathfrak{sl}(2,\mathbb{C})} \simeq \left(\mathfrak{su}(2)\right)_\mathbb{C}.
\ee
\begin{proof}
The isomorphism is guaranteed by choosing the right basis for $\mathfrak{su}(2)_\mathbb{C}$. Using matrices defined in Eq. (\ref{beta}), this is achieved with the following basis
\begin{equation*}
    \xi_1 = -2i\beta_3 = H, \quad \xi_2 = -i(\beta_1 + i\beta_2) = X, \quad \xi_3 = -i(\beta_1 - i\beta_2) = Y.
\end{equation*}
This proves the isomorphism of complex vector spaces. The extension of the $\mathfrak{su}(2)$ Lie bracket to $\mathfrak{su}(2)_\mathbb{C}$ is the Lie bracket of $\mathfrak{sl}(2,\mathbb{C})$, which ensures the Lie algebras isomorphism. 
\end{proof}
\end{lemma}

We stress that the complex vector space isomorphism can be found through direct decomposition. Specifically, we can write any element $X$ of ${\mathfrak{sl}(2,\mathbb{C})}$ as :
$$
X = \left(\frac{1}{2}(X - X^\dagger)\right) + i \left(\frac{1}{2i}(X + X^\dagger)\right),
$$
where $\left(\frac{1}{2}(X - X^\dagger)\right)$ and $\left(\frac{1}{2i}(X + X^\dagger)\right)$ are two elements of $\left(\mathfrak{su}(2)\right)$. This is to be compared with the map $\Theta^{-1}$ defined in the proof of Theorem (\ref{theorem:restrictionExtension}).

From Eq. (\ref{sl2a}) and  (\ref{sl2b}), we obtain
\be
\label{sl2c}
 \left(\mathfrak{sl}(2,\mathbb{C})^\mathbb{R}\right)_\mathbb{C} \simeq \left(\mathfrak{su}(2)\right)_\mathbb{C} \times \left(\mathfrak{su}(2)\right)_\mathbb{C}.
\ee
We get 
that the irreducible representations of finite dimensions of $\left(\mathfrak{sl}(2,\mathbb{C})^\mathbb{R}\right)_\mathbb{C}$ 
are given by two copies of the irreducible representations (of finite dimension) of $\left(\mathfrak{su}(2)\right)_\mathbb{C}$, thus characterized by two half-integers $(j_1,j_2)$.
Spaces of representation can then be written as 
\be
\label{vv}
V^{(j_1,j_2)}= \{\ket{j_1,j_2;m_1,m_2}, m_1=-j_1,-j_1+1,...,j_1-1,j_1,m_2=-j_2,-j_2+1,...,j_2-1,j_2\},
\ee
with $j_1, j_2 \in \frac{1}{2}\mathbb{N}$.

Using Lemma (\ref{theorem:so13sl2C}), we obtain that irreducible representations of finite dimensions of $\mathfrak{so}(1,3)_\mathbb{C}$, and thus those of $\mathfrak{so}(1,3)$,
 are characterized by two half-integers $(j_1,j_2)$, with 
spaces of representation $V^{(j_1,j_2)}$ (Eq. (\ref{vv})) of dimension $(2j_1+1)(2j_2+1)$.
Let us note that, because the proper Lorentz group is not compact, its non trivial irreducible representations of finite dimension are non unitarizable.

Let us see this result in a different way. As discussed in the Appendix \ref{appendix5}, $\mathfrak{so}(1,3)$ is spanned
by the three generators for the Lorentz boosts $K_{1,2,3}$
and the three generators for rotations in space $\eta_{1,2,3}$, see Eq.  (\ref{lorentzbasis2}), with the Lie brackets
\be
\label{lieso}
[\eta_i,\eta_j] = \sum_k \epsilon_{ijk} \eta_k \ \ , \ \ [K_i,K_j] = -\sum_k \epsilon_{ijk} \eta_k \ \ , \ \ [K_i,\eta_j] = \sum_k \epsilon_{ijk} K_k.
\ee
Consider the complexification of $\mathfrak{so}(1,3)$, namely $\left(\mathfrak{so}(1,3)\right)_\mathbb{C}$, it is possible to make the following combinations of generators
$$
\eta_i^{(+)}=\frac{1}{2}(\eta_i+i K_i) \ \ , \ \ \eta_i^{(-)}=\frac{1}{2}(\eta_i-i K_i),
$$
whose Lie brackets are found to be, 
$$
[\eta_i^{(+)},\eta_j^{(-)}] =0 \ \ , \ \ [\eta_i^{(+)},\eta_j^{(+)}]= \sum_k \epsilon_{ijk} \eta_k^{(+)} \ \ , \ \ [\eta_i^{(-)},\eta_j^{(-)}]= \sum_k \epsilon_{ijk} \eta_k^{(-)}.
$$
From these relations, we see that $\left(\mathfrak{so}(1,3)\right)_\mathbb{C}$ decomposes in two commuting subalgebras
of $\left(\mathfrak{su}(2)\right)_\mathbb{C}$, or
$$
\left(\mathfrak{so}(1,3)\right)_\mathbb{C} \simeq
 \left(\mathfrak{su}(2)\right)_\mathbb{C} \times \left(\mathfrak{su}(2)\right)_\mathbb{C}.
 $$
Here, operators $\eta_i^{(+)}$ are the generators for one copy $\left(\mathfrak{su}(2)\right)_\mathbb{C}$ and $\eta_i^{(-)}$ for the second one.
The relation $[\eta_i^{(+)},\eta_j^{(-)}] =0$ means that they are independent.
Then, any element of the proper Lorentz group $SO^0(1,3)$ close to the identity can be written as
\be
\label{d11}
\Lambda = e^{\sum_i \theta_i \eta_i + \gamma_i K_i} = e^{{\vec \theta}.{\vec \eta} + {\vec \gamma}.{\vec K} },
\ee
where $(\theta_i)_{i=1,2,3}$ and $(\gamma_i)_{i=1,2,3}$  are six real parameters, which also reads
\be
\Lambda =  e^{ ({\vec \theta}-i{\vec \gamma}).{\vec \eta^{(+)}} + ({\vec \theta}+i{\vec \gamma}).{\vec \eta^{(-)}} },
\ee
or
\be
\label{l2}
\Lambda =  e^{ -(i{\vec \theta}+{\vec \gamma}).(i{\vec \eta^{(+)}}) - (i{\vec \theta}-{\vec \gamma}).(i{\vec \eta^{(-)}}) }.
\ee
One should keep in mind that strictly speaking, not every element of $SO^0(1,3)$ can be written as in Eq. (\ref{l2}). This is because $SO^0(1,3)$ is not simply connected, and the exponential map is not surjective.

Let us point out that $i{\vec \eta^{(+)}}$ and $i{\vec \eta^{(-)}}$ are hermitian operators, which is why Eq. (\ref{l2}) is preferred in physics theory, where quantum observables are represented by hermitian operators.
In particular, in the irreducible representation labels $(j_1,j_2)$ above, we have
$$
\left( i{\vec \eta^{(+)}} \right)^2 {\ket{j_1,j_2;m_1,m_2}} = j_1(j_1+1) {\ket{j_1,j_2;m_1,m_2}},
$$
and
$$
\left( i{\vec \eta^{(-)}} \right)^2 {\ket{j_1,j_2;m_1,m_2}} = j_2(j_2+1) {\ket{j_1,j_2;m_1,m_2}}.
$$
At this point, we can
make a short list of the various representations of the proper Lorentz group as pairs of half-integers $(j_1,j_2)$
and then discuss for some of them what are the associated transformations, using Eq. (\ref{l2}).
\begin{itemize}
\item[$\circ$] $(0,0)$ of dimension $1$ is called the scalar representation,
\item[$\circ$] $(\frac{1}{2},0)$ of dimension $2$ is called the left spinor representation,
\item[$\circ$] $(0,\frac{1}{2})$ of dimension $2$ is called the right spinor representation,
\item[$\circ$] $(\frac{1}{2},\frac{1}{2})$ of dimension $4$ is called the vector representation,
\item[$\circ$] $(\frac{1}{2},0) \oplus (0,\frac{1}{2})$ of dimension $4$ is called the Dirac spinor representation.
\end{itemize}

Let us pause here to summarize the way we managed to determine the irreducible representations of the Lorentz group. \begin{itemize}
    \item [1. ] The Lorentz group is generated by the algebra $\mathfrak{so}(1,3)$, which consists of the generators of rotations and boosts.
    \item [2. ] The complexification of $\mathfrak{so}(1,3)$ is the Lie algebra $\left(\mathfrak{sl}(2,\mathbb{C})^\mathbb{R}\right)^\mathbb{C}$. It suffices to determine the irreducible representations of $\left(\mathfrak{sl}(2,\mathbb{C})^\mathbb{R}\right)^\mathbb{C}$ to get those of $\mathfrak{so}(1,3)$.
    \item [3. ] $\left(\mathfrak{sl}(2,\mathbb{C})^\mathbb{R}\right)^\mathbb{C} = \mathfrak{sl}(2,\mathbb{C}) \times \mathfrak{sl}(2,\mathbb{C})$ so a representation of $\mathfrak{so}(1,3)$ is given by two representations of $\mathfrak{sl}(2,\mathbb{C})$.
    \item [4. ] $\mathfrak{sl}(2,\mathbb{C}) \simeq \mathfrak{su}(2)_\mathbb{C}$ whose representations are labelled by a single half integer $j$
    \item [5. ] The irreducible representations of $\mathfrak{so}(1,3)$ are labeled by pairs of half-integers $(j_1,j_2)$, corresponding to the two $\mathfrak{su}(2)$ factors.
\end{itemize}

The Lorentz transformations (see Eq. (\ref{l2})) belonging to the $(\frac{1}{2},0)$ representation (of dimension $2$) can be written as
$$
\Lambda_L({\vec \theta},{\vec \gamma}) =  e^{ -(i{\vec \theta}+{\vec \gamma}).(i{\vec \eta^{(+)}}) }.
$$
Moreover, we can verify that
$ i{\vec \eta^{(+)}} = \frac{\vec \sigma}{2}$,
where ${\vec \sigma}$ are the Pauli $2 \times 2$ complex matrices. We get
$$
\Lambda_L({\vec \theta},{\vec \gamma}) =  e^{ -\frac{1}{2}(i{\vec \theta}+{\vec \gamma}).{\vec \sigma} }.
$$
These transformations act on  physical fields (operators of the theory) which are necessarily of the form.
$$
\psi_L(x) = \begin{pmatrix}
A  \\
B 
\end{pmatrix}(x) \in \mathbb{C}^2.
$$
See appendix 6 for a more formal discussion of this last point.
The transformation of $\psi_L$ reads
$$
\psi'_L(x') = \Lambda_L({\vec \theta},{\vec \gamma}) \psi_L(x),
$$
or
$$
\psi'_L(x) = \Lambda_L({\vec \theta},{\vec \gamma}) \psi_L(\Lambda^{-1}({\vec \theta},{\vec \gamma}).x).
$$

For the Lorentz transformations belonging to the $(0,\frac{1}{2})$, we have similarly
$$
\Lambda_R({\vec \theta},{\vec \gamma}) =  e^{ -\frac{1}{2}(-{\vec \theta}+i{\vec \gamma}).{\vec \sigma} }.
$$
They act on  physical fields of the form
$$
\psi_R = \begin{pmatrix}
A'  \\
B' 
\end{pmatrix} \in \mathbb{C}^2.
$$ 
The Lorentz transformation of $\psi_R$ reads
$$
\psi'_R(x) = \Lambda_R({\vec \theta},{\vec \gamma}) \psi_R(\Lambda^{-1}({\vec \theta},{\vec \gamma}).x).
$$

We recall that, as the proper Lorentz group is not compact, $\Lambda_{L,R}$ are non unitary matrices
$$
\Lambda_{L,R}^\dagger \ne \Lambda_{L,R}^{-1}.
$$
However, we can verify  the interesting property 
$$
\Lambda_L = (\Lambda_R^\dagger)^{-1} \ \ , \ \ \Lambda_R = (\Lambda_L^\dagger)^{-1}.
$$

Now, if we consider the representation $(\frac{1}{2},0) \oplus (0,\frac{1}{2})$, it is easy to see that the Lorentz transformations can be decomposed as
$$
\Lambda_D = \Lambda_L \oplus \Lambda_R = 
\begin{pmatrix}
\Lambda_L({\vec \theta},{\vec \gamma}) & 0 \\
0 & \Lambda_R({\vec \theta},{\vec \gamma}) 
\end{pmatrix}
=
\begin{pmatrix}
\Lambda_L({\vec \theta},{\vec \gamma}) & 0 \\
0 & (\Lambda_L^\dagger({\vec \theta},{\vec \gamma}))^{-1}.
\end{pmatrix}
$$
With fields Dirac spinors, 
$$
\psi_D(x) = 
\begin{pmatrix}
\psi_L(x)  \\
\psi_R(x)
\end{pmatrix} \in \mathbb{C}^4,
$$
we have
$$
\psi'_D(x) = \Lambda_D({\vec \theta},{\vec \gamma}) \psi_D(\Lambda^{-1}({\vec \theta},{\vec \gamma}).x).
$$
$\psi_D$ is a fundamental field in physics that  describe matter particles.
The Dirac bispinor field is the fundamental object that describes fermionic particles, that is matter. The question of why it is used rather than the irreducible fields that are the left and right handed spinor representations is a subtle one. It goes back to the complete structure of $SO(1,3)$, which included 4 connected components, taking into account parity and time reversal transformations. The Dirac bispinor representation can be extended to a representation of the full Lorentz group $SO(1,3)$, which is not the case for the left and right spinor representations. In words, it means that contrary to the right and left handed spinors, the Dirac spinor representation allows to tackle questions related to parity and time reversal. Additionnally, it should be noted that mass terms in particle physics come from the mixing between left and right handed components of the Dirac spinor. 


\section{Conclusion}

In physics theory, the kind of particles that are admissible in the universe, provided that the Lorentz symmetry is ensured,
is based on the irreducible representations of finite dimension of the Lie algebra $\mathfrak{so}(1,3)$.
In order to derive those algebraic structures, it is necessary to consider the complexification of $\mathfrak{so}(1,3)$,
namely $\mathfrak{so}(1,3)_\mathbb{C}$.
The process of complexification is therefore of essential use in physics. In this paper, we have defined it precisely with different views,
see Definition (\ref{definition:complexification}). Also, this definition is accompanied with an important result, see Theorem (\ref{theorem:restrictionExtension})
$$
    \left(\mathfrak{g}^\mathbb{R}\right)_\mathbb{C} \simeq \mathfrak{g} \times \bar{\mathfrak{g}},
$$
that we prove step by step in the core of the paper.
This result is needed to obtain the relation
$$
 \mathfrak{so}(1,3)_\mathbb{C} \simeq \left(\mathfrak{su}(2)\right)_\mathbb{C} \times \left(\mathfrak{su}(2)\right)_\mathbb{C},
$$
from which irreducible representations (of finite dimensions) of the proper Lorentz group can be obtained.
At this point, we have been able to identify the mathematical structure of physical objects that are admissible in the universe as a consequence
of Definition (\ref{definition:complexification}) and Theorem (\ref{theorem:restrictionExtension}).
In particular, we have shown that finite representations of the proper Lorentz group are characterized by pairs of half-integers $(j_1,j_2)$, which
determine  unambiguously the physical object associated to the given representation. 
For example, the representation $(\frac{1}{2},0) \oplus (0,\frac{1}{2})$ of dimension $4$ is called the Dirac spinor representation, it corresponds to  
matter particles called fermions.

\section{Appendix, proof of Lemma (\ref{theorem:so13sl2C})}
\label{appendix5}

$$
\mathfrak{so}(1,3)_\mathbb{C} \simeq \left(\mathfrak{sl}(2,\mathbb{C})^\mathbb{R}\right)_\mathbb{C}
$$
\begin{proof}
There are several ways to prove this result.
One way is to show first that 
\be
\label{pr1}
\mathfrak{so}(1,3)_{\mathbb{C}} \simeq \mathfrak{so}(4)_{\mathbb{C}},
\ee
with
\be
\label{pr2}
\mathfrak{so}(4)_{\mathbb{C}} \simeq  \mathfrak{su}(2)_{\mathbb{C}} \times \mathfrak{su}(2)_{\mathbb{C}}.
\ee
From these two relations, this is then immediate that
$$
\mathfrak{so}(1,3)_{\mathbb{C}} \simeq \left(\mathfrak{sl}(2,\mathbb{C})^\mathbb{R}\right)_\mathbb{C}.
$$

Let us give the main lines of the proof of Eq. (\ref{pr1}) first. 
A basis of the Lie algebra $\mathfrak{so}(1,3)$ is given by the three generators of the Lorentz boosts $K_{1,2,3}$
and the three generators of rotations in space $\eta_{1,2,3}$, namely
\be
\label{lorentzbasis1}
K_1 =
\begin{pmatrix}
0 & 1 & 0 & 0  \\
1 & 0 & 0 & 0 \\
0 & 0 & 0 & 0 \\
0 & 0 & 0 & 0 
\end{pmatrix}
\ \ \
K_2 =
\begin{pmatrix}
0 & 0 & 1 & 0  \\
0 & 0 & 0 & 0 \\
1 & 0 & 0 & 0 \\
0 & 0 & 0 & 0 
\end{pmatrix}
\ \ \
K_3 =
\begin{pmatrix}
0 & 0 &  0 & 1  \\
0 & 0 & 0 & 0 \\
0 & 0 & 0 & 0 \\
1 & 0 & 0 & 0 
\end{pmatrix}
$$
$$
\eta_1 =
\begin{pmatrix}
0 & 0 & 0 & 0  \\
0 & 0 & 0 & 0 \\
0 & 0 & 0 & -1 \\
0 & 0 & 1 & 0 
\end{pmatrix}
\ \ \
\eta_2 =
\begin{pmatrix}
0 & 0 & 0 & 0  \\
0 & 0 & 0 & 1 \\
0 & 0 & 0 & 0 \\
0 & -1& 0 & 0 
\end{pmatrix}
\ \ \
\eta_3 =
\begin{pmatrix}
0 & 0 & 0 & 0  \\
0 & 0 & -1 & 0 \\
0 & 1 & 0 & 0 \\
0 & 0 & 0 & 0 
\end{pmatrix}
\ee
Let us write
\be
\label{lorentzbasis2}
\mathfrak{so}(1,3) = \{K_1,K_2,K_3,\eta_1,\eta_2,\eta_3\},
\ee
which means that the (real) Lie algebra $\mathfrak{so}(1,3)$ is spanned by the generators $\{K_1,K_2,K_3,\eta_1,\eta_2,\eta_3\}$. Also, we can check
that the Lie brackets read
\be
\label{tz1}
[\eta_i,\eta_j] = \sum_k \epsilon_{ijk} \eta_k \ \ , \ \ [K_i,K_j] = -\sum_k \epsilon_{ijk} \eta_k \ \ , \ \ [K_i,\eta_j] = \sum_k \epsilon_{ijk} K_k.
\ee
Let us now consider the (real) Lie algebra $\mathfrak{so}(4)$. There are $6$ generators, corresponding to the $6$
possible rotations in $4$ dimensions. Let us write 
$$
\mathfrak{so}(4) = \{A_1,A_2,A_3,B_1,B_2,B_3\}.
$$
Also, we can compute easily the Lie brackets
$$
[B_i,B_j] = \sum_k \epsilon_{ijk} B_k \ \ , \ \ [A_i,A_j] = \sum_k \epsilon_{ijk} B_k \ \ , \ \ [A_i,B_j] = \sum_k \epsilon_{ijk} A_k.
$$
We can remark that if we make the transformation $\tilde{A}_j= i A_j$, we get
\be
\label{tz2}
[B_i,B_j] = \sum_k \epsilon_{ijk} B_k \ \ , \ \ [\tilde{A}_i,\tilde{A}_j] = -\sum_k \epsilon_{ijk} B_k \ \ , \ \ [\tilde{A}_i,B_j] = \sum_k \epsilon_{ijk} \tilde{A}_k,
\ee
which is the same Lie algebra structure as in Eq. (\ref{tz1}). 
The point here is that by making the transformation $\tilde{A}_j= i A_j$, we subsequently consider a new Lie algebra. Following Definition (\ref{definition:complexification}), 
this corresponds to  the complexified 
Lie algebra $\left(\mathfrak{so}(4)\right)_\mathbb{C}$. The correspondance between the two Lie algebra given by the Lie brackets (\ref{tz1}) and (\ref{tz2})
is  correct but only for the complexified Lie algebra.
This proves Eq. (\ref{pr1}).

The proof of Eq. (\ref{pr2}) is similar. We start from the Lie algebra of $\mathfrak{so}(4)$ and make the transformation.
$$
B^{(+)}=\frac{1}{2}(B_i+A_i) \ \ , \ \ B_i^{(-)}=\frac{1}{2}(B_i-A_i).
$$
This gives
$$
[B_i^{(+)},B_j^{(-)}] =0 \ \ , \ \ [B_i^{(+)},B_j^{(+)}]= \sum_k \epsilon_{ijk} B_k^{(+)} \ \ , \ \ [B_i^{(-)},B_j^{(-)}]= \sum_k \epsilon_{ijk} B_k^{(-)}.
$$
Then
$$
\mathfrak{so}(4) \simeq  \mathfrak{su}(2) \times \mathfrak{su}(2),
$$
which implies Eq. (\ref{pr2}) after complexification.

\end{proof}

\section{Appendix : Poincaré group and fields}
\label{appendix6}
Since this discussion is oriented towards its applications in physics, let us get a feeling of how this mathematical framework translates into particle physics.

The symmetry group of the Universe in particle physics is the Poincaré group $SO^0(1,3)\ltimes \mathbb{R}^4$, whose composition law is given by, $$
(\Lambda_1, a_1)\cdot (\Lambda_2, a_2) = (\Lambda_1 \Lambda_2, a_1 + \Lambda_1 a_2).
$$
On physical states these acts through a unitary representation, $U(\Lambda,a)$. This means that if $\ket{\phi}$ is a physical state belonging to the Hilbert space $\mathcal{H}$ of the theory, then, \begin{equation*}
    \ket{\phi}\overset{(\Lambda,a)}{\longrightarrow} U(\Lambda,a) \ket{\phi}.
\end{equation*}
Picking a passive point of view for the symmetry transformations, the operators are transformed rather than the states. Namely, if $O$ is an operator, we have
$$
O \overset{(\Lambda,a)}{\longrightarrow} U(\Lambda,a)^\dagger O U(\Lambda,a).
$$

The Poincaré group is non compact so that there are no non-trivial finite dimensional unitary representations. Consequently, $\mathcal{H}$ is infinite dimensional, and the operators needs to be thought of as continuous acting on a position or momentum basis. Operators becomes maps from the Minkowski spacetime $\mathbb{R}^{1,3}$ to the endomorphisms of the Hilbert space, $\mathcal{L}(\mathcal{H})$. The transformation rules for such operators are not obvious and cannot be derived from first principles or Wigner's classification. The transformation rules are rather axiomatic and define \emph{quantum fields} \cite{weinberg}. 

\begin{definition}
    A quantum field is an operator valued distribution $O$ of finite dimension (that has a finite number of components) that transforms as, \begin{equation}
        \label{eq:field_transformation}
        U(\Lambda,a)^\dagger O(x) U(\Lambda,a) = D(\Lambda) O(\Lambda^{-1}(x-a)),
    \end{equation}
    where $D(\Lambda)$ is a finite dimensional representation of the Lorentz group.
\end{definition}

There is still some intuition to unpack with this axiomatic definition. Let us highlight how the operators transform under pure translations. Such transformations corresponds to the subset $$
\left\{ (1, a)\;|\; a\in\mathbb{R}^4\right\}\subset SO(1,3)\ltimes \mathbb{R}^4$$
of the Poincaré group. For such transformations, we have, \begin{equation*}
    U(1, a)^\dagger O(x) U(1, a) = D(1)O(x-a) = O(x-a).
\end{equation*}
Think about such an operator acting on a state. The above equation stresses that the translation can be realized in 2 ways, either translate the state by the quantity $+a$ and apply the operator, left unchanged, or translate the operator by the quantity $-a$ and apply it to the unchanged state. This is the essence of the difference between active (acting on the states) and passive (acting on the operators) transformations. 

Formally, this means that the space of operators can be thought of as the tensor product $\mathcal{L}(\mathcal{H}) \otimes V$, where $V$ is a finite dimensional vector space on which the representation $D(\Lambda)$ acts. The matrix $D$ really is a representation of the Lorentz group on $V$, a finite dimensional complex vector space, that we may decompose into irreducible representations. This way, we can arrange the fields into \emph{irreducible fields}, that are fields for which the matrix $D$ furnishes an irreducible representation of the Lorentz group. For irreducible fields, the matrix $D$ is fully determined by our previous work (section 3) and characterized by the two half integers $(j_1,j_2)$ introduced in section 3.
Then, in an $n=(2j_1+1)(2j_2+1)$ representation, $D$ is an $n \times n$ matrix written as $\Lambda$ in Eq. (\ref{d11}). We have
$$
D = e^{{\vec \theta}.{\vec \eta} + {\vec \gamma}.{\vec K} },
$$
where $(\theta_i)_{i=1,2,3}$ et $(\gamma_i)_{i=1,2,3}$  are six real parameters and ${\vec \eta}$, ${\vec K}$ also $n \times n$ matrices, satisfying the Lie algebra structure of
Eq. (\ref{lieso}). For example,
let us recall (see section 3) the left and right spinor representations for which the $D$ matrices are given by
$$
    \Lambda_L({\vec \theta},{\vec \gamma})=e^{ -(i{\vec \theta}+{\vec \gamma}).(i{\vec \eta^{(+)}}) } \;\text{for}\;(\frac{1}{2},0) \quad \text{and} \quad \Lambda_R({\vec \theta},{\vec \gamma})=  e^{ -(i{\vec \theta}-{\vec \gamma}).(i{\vec \eta^{(-)}}) } \;\text{for}\;(0,\frac{1}{2}),
$$
Both these representations have dimension $2$ so that the associated fields must be of dimension $2$, 
$$
\Psi_L(x) = \begin{pmatrix}
    \psi_{L1}(x) \\
    \psi_{L2}(x)
\end{pmatrix} \quad \text{and} \quad \Psi_R(x) = \begin{pmatrix}
    \psi_{R1}(x) \\
    \psi_{R2}(x)
\end{pmatrix}.
$$
The equality of dimension is not sufficient to declare these are the same. In fact, they definitely correspond to different physical objects. 

At this point, we can complete the short list of the various representations of the proper Lorentz group as pairs of half-integers $(j_1,j_2)$ (section 3)
with a reference to the physical objects described.
\begin{itemize}
\item[$\circ$] $(0,0)$ of dimension $1$ is called the scalar representation, it corresponds to the Higgs field
\item[$\circ$] $(\frac{1}{2},0)$ of dimension $2$ is called the left spinor representation, it corresponds to the left-handed neutrino
\item[$\circ$] $(0,\frac{1}{2})$ of dimension $2$ is called the right spinor representation, it corresponds to the right-handed anti-neutrino
\item[$\circ$] $(\frac{1}{2},\frac{1}{2})$ of dimension $4$ is called the vector representation, it corresponds to the gauge bosons
\item[$\circ$] $(\frac{1}{2},0) \oplus (0,\frac{1}{2})$ of dimension $4$ is called the Dirac spinor representation, it corresponds to the Dirac fermions.
\item[$\circ$] $(1,1)$ of dimension $9$ is called the tensor representation, it corresponds to the graviton.
\end{itemize}

{\bf Data availability} \\
No data was used for the research described in the article.

{\bf Declaration of interests} \\
The authors declare that they have no known competing financial interests or personal relationships that could have appeared to influence the work reported in this paper.


\end{document}